\documentclass{jfm}
\usepackage{graphicx,ulem}
\usepackage{epstopdf, epsfig}
\usepackage{amsmath}
\usepackage{amsfonts}
\usepackage{amssymb}

\usepackage{xcolor}

\newcommand{\diff}{\text{d}}

\shorttitle{Vortex cluster arising from an axisymmetric inertial wave attractor}
\shortauthor{S. Boury et al.}

\title{Vortex cluster arising from an axisymmetric inertial wave attractor}

\author{S. Boury\aff{1}
  \corresp{\email{samuel.boury@ens-lyon.fr}}, 
  I. Sibgatullin\aff{2,3},
  E. Ermanyuk\aff{4},
  N. Shmakova\aff{4},
  P. Odier\aff{1},
  S. Joubaud\aff{1,5},
  L.R.M. Maas\aff{6}
  \and T. Dauxois\aff{1}}

\affiliation{\aff{1}Univ Lyon, ENS de Lyon, Univ Claude Bernard, CNRS, Laboratoire de Physique, F-69342 Lyon, France
\aff{4}Lavrentyev Institute of Hydrodynamics, av. Lavrentyev 15, Novosibirsk, 630090, Russia
\aff{2}P.P. Shirshov Institute of Oceanology, Nahimovskiy prospect 36, Moscow, 117997, Russia
\aff{5}Institut Universitaire de France (IUF), 1 rue Descartes 75005 Paris, France
\aff{6}Institute for Marine and Atmospheric Research, Utrecht University, 3584 CC Utrecht, The Netherlands
\aff{3}Ivannikov Institute for System Programming, str. Alexander Solzhenitsyn 25, Moscow, 109004, Russia}

\begin{document}

\maketitle

\begin{abstract}
	

We present an experimental study of the saturated non-linear dynamics of an inertial wave attractor in an axisymmetric geometrical setting. The experiments are carried out in a rotating ring-shaped fluid domain delimited by two vertical coaxial cylinders, a conical bottom, and a horizontal deformable upper lid as wave generator: the meridional cross-section of the fluid volume is a trapezium, while the horizontal cross-section is a ring. First, the fluid is set into a rigid-body rotation. Thereafter, forcing is introduced into the system via axisymmetric low-amplitude volume-conserving oscillatory motion of the upper lid. After a short transient of about $10$ forcing periods, a quasi-linear regime is established, with an axisymmetric inertial wave attractor. The attractor is prone to instability: at long time-scale (order 100 forcing periods) a saturated fully non-linear regime develops as a consequence of an energy cascade draining energy towards a slow two-dimensional manifold represented by a regular polygonal system of axially-oriented cyclonic vortices that are slowly precessing around the inner cylinder. We show that this slow two-dimensional manifold manifests a persistent slow prograde motion and a strong cyclonic-anticyclonic asymmetry quantified by the time-evolution of the probability density function of the vertical vorticity. 

\end{abstract}

\begin{keywords}
Authors should not enter keywords on the manuscript, as these must be chosen by the author during the online submission process and will then be added during the typesetting process (see http://journals.cambridge.org/data/\linebreak[3]relatedlink/jfm-\linebreak[3]keywords.pdf for the full list)
\end{keywords}

	\section{Introduction}

	Energy cascade in rotating fluids has received significant attention due to its relevance to geo- and astrophysical fluid dynamics and due to the rich complexity of the non-linear multi-scale interplay between coherent vortical structures, inertial waves and background small-scale nearly isotropic turbulence \citep{Greenspan1968,HopfingervanHeijst1993,Davidson2013,GodeferdMoisy2015}. Inertial waves supported by rotating fluids, with the Coriolis force acting as restoring force, represent an essential ingredient of the cascade. The crucial role of inertial waves is assured by [i] the specific form of the dispersion relation, which contains no length scale, and [ii] the possibility of a cascade of wave-wave interactions due to non-linear terms in the Navier-Stokes equations governing the dynamics of rotating fluids.
	
	The dispersion relation of inertial waves obtained by seeking plane-wave solutions of the linearized inviscid Navier-Stokes equations reads $\omega=f k_{z}/k=f \cos \beta$, where $\omega$ is the wave frequency, $f=2\Omega$ is the Coriolis parameter with $\Omega$ the rate of the background rigid-body rotation of the fluid, and $k_{z}$ (respectively $k$) is the vertical component (respectively magnitude) of the wave vector $\mathbf{k}$ inclined at angle $\beta$ to the vertical $z$-axis, which is taken as the axis of rotation. A similar type of dispersion relation $\omega=N k_h/k=N \sin \beta$ holds for internal waves, with the buoyancy frequency~$N$ replacing $f$, and $k_{h}$ the horizontal wave vector. The absence of any length scale in the dispersion relation for inertial and internal waves implies that the global large-scale wave pattern depends on the geometry of wave generators and on the geometry that delimits the fluid volume -- in particular, for the ocean, its bathymetry. Therefore, a rich variety of wave motions is encountered in rotating and stratified fluids as identified in early pioneering studies, and explored in detail in the subsequent literature: normal modes in bounded domains of simple geometry (sphere, axial cylinder, slanted rectangular box) \citep{AldridgeToomre1969,McEwan1970,McEwan1971, maas2003, bewley2007, lamriben2011, boisson2012, wu2020a}, wave beams emanating from isolated oscillatory sources \citep{Gortler1943,Hendershott1969,ThomasStevenson1972,MowbrayRarity1967}, and webs of wave beams (wave attractors) in bounded domains with sloping walls \citep{Stern1963,Bretherton1964,Stewartson1971,Stewartson1972,MaasLam1995,Maasetal1997,MandersMaas2003}. Of particular interest is the latter configuration in the context of the present paper.
	
	Due to the form of the internal and inertial wave dispersion relations, wave reflection on a solid boundary follows a very specific law and is, in general, non-specular \citep{Phillips1963,Eriksen1982,MandersMaas2004,Maas2005}. To be more specific, in two-dimensional domains, this law leads to a focusing or a defocusing effect of wave beams upon reflection at sloping walls \citep{dauxois1999}. In bounded or quasi-bounded two-dimensional domains, focusing prevails: the iterative process of subsequent wave reflections leads to the formation of a limit cycle, called a wave attractor, where the wave energy is concentrated \citep{MaasLam1995,Maasetal1997}. Relevant to the topic of the present study  is the first experimental observation of an inertial wave attractor in an elongated trapezoidal tank that showed the generation of a persistent mean flow, right above the location where the attractor was being focused over the sloping bottom. This mean flow was speculated to be the result of the breaking of focused inertial waves, leading to the mixing of the background radial stratification in angular momentum with which the solidly-rotating, homogeneous-density fluid is endowed \citep{maas2001}. In a three-dimensional setting, the variety of possible configurations is significantly enriched, involving the possibility of wave-energy trapping on a limit cycle located at certain preferential planes of motion provided that there is a billiard pathway connecting this plane and the initial direction of the wave-energy propagation \citep{Hazewinkeletal2011,Pilletetal2018}. The inertial-wave-ray billiard corresponding to the geophysically important case of a rotating spherical shell favors the formation of an attractor in the meridional plane \citep{Bretherton1964, Stewartson1972, friedlander1982, MaasHarlander2007,RabitiMaas2013}. Accordingly, the rich literature on the linear dynamics of inertial wave attractors in rotating spherical layers considers the motions in the ring-shaped meridional slices \citep{friedlander1982, dintrans1999, RieutordetalPRL2000,Rieutordetal2001,RieutordValdettaro2010} and disregards the azimuthal coordinate.
	
	It is noteworthy that the purely geometrical mechanism of iterative focusing, which is linear, is at the origin of a spectacular forward energy cascade in wave attractors: the energy injected into the system at global scale (i.e. at the scale of the system itself) is transferred to the scale corresponding to the width of the attractor branches, which, even in laboratory experiments, can be an order of magnitude smaller than the global scale \citep{BrouzetetalPRF2017}. This small scale, or width of the inertial and internal wave beams in the linear regime, is set by the balance between geometric focusing and viscous dissipation and can be theoretically predicted with good agreement to experimental observations \citep{Rieutordetal2001,HBDM2008,Grisouardetal2008}. Other wave-damping mechanisms such as interaction of waves with convective motions, ohmic damping in presence of magnetic field in conductive
fluids, and non-linear parametric decay into secondary waves of shorter wavelength have also been proposed by \cite{Ogilvie2005}, where a generic case with a weak non-viscous ``frictional'' damping force has been considered. Further, it has been shown experimentally that at sufficiently high level of injected energy internal wave attractors are prone to Triadic Resonance Instability (TRI) \citep{Scolanetal2013}. The replacement of purely viscous damping by the flux of energy carried by small-scale secondary waves (generated via TRI) away from the primary waves (i.e. from the beams of attractor) introduces a new non-linear scaling for the beam width \citep{BrouzetetalPRF2017}.  Similar effects in inertial waves have been observed in numerical simulations \citep{JouveOgilvie2014}. Let us note in passing that \cite{JouveOgilvie2014} considered a two-dimensional setting, physically corresponding to a torus of infinite radius having a tilted-square ``meridional'' cross-section, so that any three-dimensional effects occurring in ``equatorial'' planes were completely excluded. The development of the energy cascade in wave attractors with the increase of injected energy leads to wave turbulence, with a significant occurrence of overturning events generating irreversible mixing \citep{BrouzetetalEPL2016,Brouzetetal2017,{Davisetal2020}}, and such a cascade reaches a statistically steady state when the balance is established between the injected and dissipated energy \citep{JouveOgilvie2014,Davisetal2019}.    
	
	It should be stressed that the overturning events and subsequent mixing are important constituents of the full energy cascade in internal wave attractors, and that they clearly fall apart from the wave turbulence formalism. Similarly, inertial wave turbulence plays an important but not exclusive role in the non-linear dynamics of inertial wave attractors described in the present paper. The full scope of dynamic events constituting the energy cascade in inertial wave attractors extends well beyond the wave-turbulence framework and should be discussed in the rich context of the literature on turbulence in rotating fluids. The wealth of this literature is such that, in this paper, we restrict ourselves to a cursory discussion of effects directly relevant to the present study. The reader interested in the current state of the art is relegated to \cite{Davidson2013,GodeferdMoisy2015}.
	
	The focus of the interest in rotating wave turbulence \citep{Davidson2013,GodeferdMoisy2015} lays at the anisotropy of scales along the directions parallel and perpendicular to the axis of rotation, the presence of direct and inverse cascades of the key dynamically important quantities (energy, enstrophy, etc.), non-linear wave-wave interactions among inertial waves, and the development of coherent vortex structures aligned with the axis of rotation. The importance of these issues has been identified in early experimental studies with grid-generated turbulence in rotating tanks and their numerical counterparts (see for example \cite{Hopfingeretal1982,GodeferdLollini1999}). 
	
	In order to study in isolation the effect of rotation on (initially isotropic) turbulence, considerable attention has been focused on theoretical investigations in domains of infinite extent and numerical simulations in triply periodic boxes (e.g. \cite{Waleffe1993,Cambonetal1997}). It has been shown that the anisotropy develops due to non-linear wave-wave interactions modified by rotation and concentrates energy in the plane normal to the rotation axis at a slow two-dimensional manifold \citep{Cambonetal1997}. The relevance of the wave turbulence formalism and results of numerical simulations in triply periodic boxes to the experimental reality involving secondary currents, wall-induced vorticity, and formation of Ekman and Stewartson boundary layers remains an open issue. Indeed, various saturated turbulent regimes ranging between quasi-two-dimensional and wave turbulence can be obtained in numerical simulations in triply periodic domains depending on the damping mechanism imposed onto the geostrophic component to mimic the interaction with rigid boundaries (see, e.g. \cite{LeReunetal2017}). Therefore, the experimental investigation of saturated turbulence regimes in rotating fluids attracts significant interest. For such studies, the choice of the range of experimental parameters and geometric setup remains a non-trivial issue (see e.g. \cite{GodeferdMoisy2015}). Typically, to ensure the development of a fully non-linear energy cascade one needs to ensure a low value of the Ekman number $E=\nu /(2\Omega L^2)$, where $\nu$ is kinematic viscosity and $L$ is the global length scale, which for the bounded fluid has the meaning of the container size. Further, the effect of rotation must be sufficiently strong and therefore the global Rossby number $Ro^{L}=U/(2\Omega L)$ (here $U$ is a velocity scale) must be sufficiently low. However, $Ro^{L}$ cannot be vanishingly small since it is responsible for triggering non-linear effects. The micro-dynamics of the emerging vortex structures can be conveniently quantified by the micro-Rossby number $Ro^{\xi_z}=\xi_{z}/(2\Omega)$, where $\xi_{z}$ is the perturbative vertical vorticity measured in the rotating frame. The skewness of the Probability Density Functions (PDFs) of micro-Rossby numbers is known to reflect the symmetry breaking of cyclonic/anticyclonic motions, which is a well-known property of rotating systems \citep{Pedley1969,Bradshaw1969,HopfingervanHeijst1993}. In many cases it is also convenient to introduce an appropriately averaged micro-Rossby number $\overline{Ro^\mathrm{RMS}}=\,\overline{\left\langle\xi_{z}\right\rangle}/(2\Omega)$, where brackets $\left\langle\cdot\right\rangle$ denote the Root Mean Square (RMS) value of a given quantity (here $\xi_{z}$) calculated over a zone of interest, and the overline denotes the temporal average over one forcing period. In particular, the appropriately averaged Rossby number $\overline{Ro^\mathrm{RMS}}$ is known to take a certain value when the three-dimensional to two-dimensional energy transfer is maximized \citep{BourouibaBartello2007}.

	The goal of our experimental investigation is to study the non-linear fate of an inertial wave attractor in an axisymmetric setting under experimental conditions which are compatible with the formation of a slow two-dimensional manifold coupled to the genuinely three-dimensional inertial wave field. This is achieved by designing a setup in the form of a (horizontal) rotating annulus having a (vertical) trapezoidal cross-section, thereby admitting a wave attractor structure in meridional planes and coherent vortex structures in equatorial plane. This work has been preceded by a preliminary numerical simulation of \cite{Sibgatullinetal2017} performed with the help of the spectral element method based on the code Nek5000 \citep{FischerRonquist1994}, which proved to be highly efficient in fully three-dimensional simulations of internal wave attractors \citep{Brouzetetal2016}. The simulation has been run at a relatively weak forcing and demonstrated the loss of axial symmetry due to the onset of TRI and gradual build-up of inertial wave turbulence, but the saturated state, corresponding to the formation of a slow two-dimensional manifold, has not been reached. In the present paper, we address precisely this intriguing and previously overlooked issue. We explore experimentally the formation of coherent vortex structures and measure the saturated value of the zonally-averaged Rossby number as well as time-evolution of PDFs of the micro-Rossby number indicative of cyclonic/anticyclonic asymmetry. Surprisingly, the observed coherent structures develop a collective self-organized behavior in the form of a polygonal pattern of cyclonic vortices. This manuscript is organised as follows. The experimental apparatus is described in section $2$. The experimental results are presented in section $3$, with the linear regime at early times (subsection $3.1$) and the non-linear regime arising afterwards (subsection $3.2$). Then, our conclusions are detailed in section $4$.

	\section{Experimental apparatus}
	
\begin{figure}
		\centering
		\includegraphics[scale=0.8]{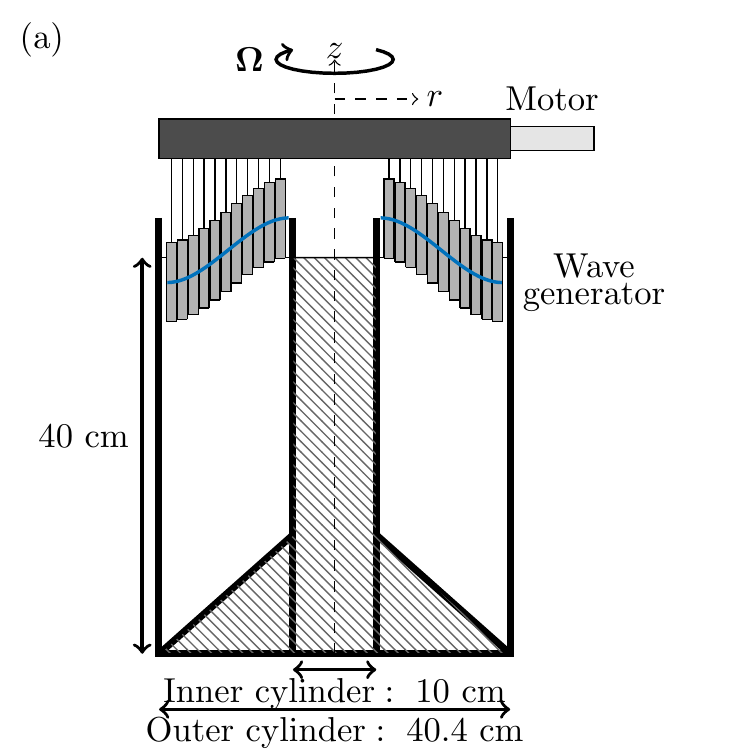}
		\includegraphics[scale=0.8]{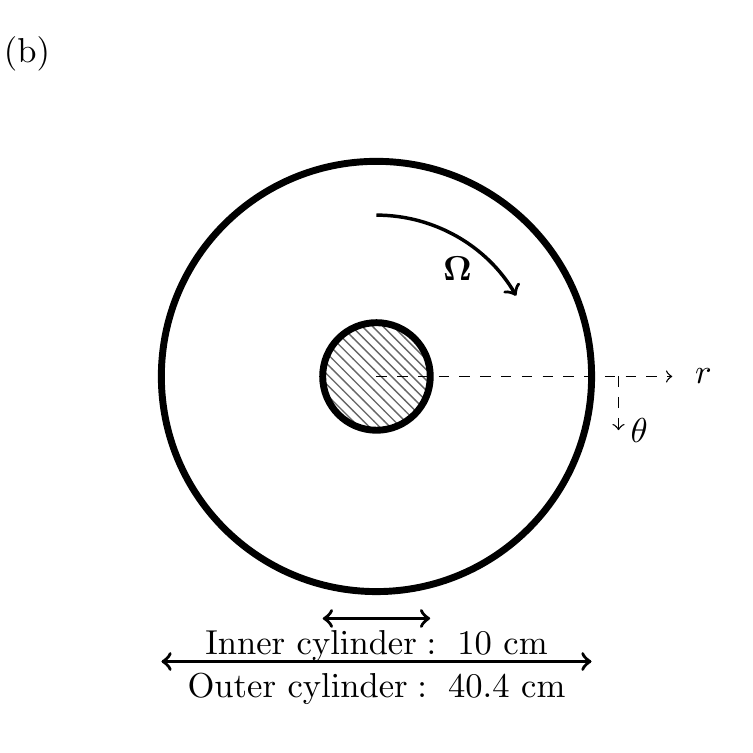}
		\caption{Schematic of the experimental apparatus in (a) a vertical cross-section, and (b) a horizontal cross-section. The hashed area is out of the experimental domain. The trapezoidal region of interest in the vertical plane is located between the inner and the outer cylinders.}
		\label{fig:Cuve}
	\end{figure}

	Figure~\ref{fig:Cuve} presents a schematic of the experimental apparatus in the vertical and horizontal cross-sections. The region of interest is bounded by two vertical acrylic cylinders, by the wave generator at the top, and by an acrylic conical surface at the bottom. The outer and inner radii of the domain are $R_1=20.2\mathrm{~cm}$ and $R_0= 5.0\mathrm{~cm}$, respectively. The generatrix of the conical bottom surface has an inclination of $45^\circ$, and the apex of the cone points upwards. Note that this cone can also be reversed upon needs: the choice taken in the present study is explained in Section 3.
	
	In a vertical (meridional) cross-section of the setup, two trapezoidal domains are facing each other as shown in figure~\ref{fig:Cuve}: note that the amplitude of the wave generator is greatly exagerated and the upper bound of the fluid domain is actually nearly flat. The depth of fluid measured along the generatrix of the outer cylinder is $H=40\mathrm{~cm}$. In a horizontal (equatorial) cross-section, the experimental domain is a ring of width $L=R_1-R_0$. The whole setup is inserted into a square acrylic tank of $100\mathrm{~cm} \times 100\mathrm{~cm}$ horizontal section and $65\mathrm{~cm}$ height, as used in \cite{boury2019}. Each part of the facility is rigidly fixed to prevent any parasitic vibration when the whole setup, mounted on the rotating table, is brought to rotation at angular velocity $\Omega=2\pi/T$ where $T$ is the rotation period. The axis of rotation of the table coincides with the symmetry axis of the setup.  
	
\begin{figure}
		\centering
		\includegraphics[width=0.9\textwidth]{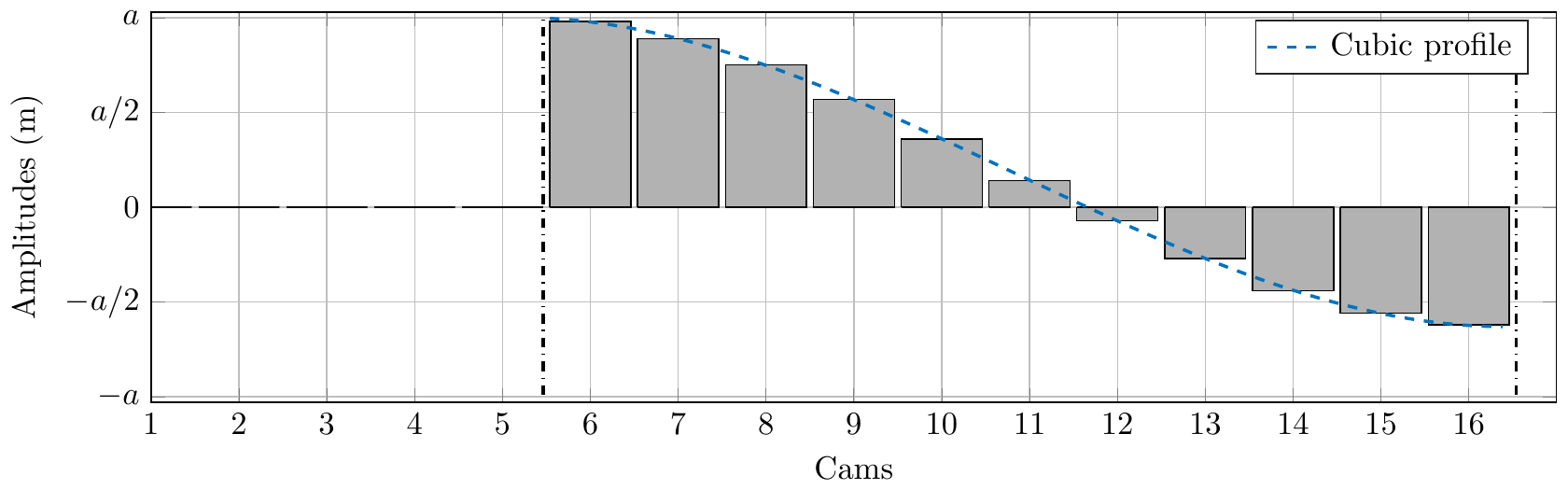}
		\caption{Configuration of the generator. The amplitude $a$ of the sixth cam is set to $2.5\mathrm{~mm}$. Out of the experimental domain, the first five cams have their amplitude set to zero. The two dash-dotted lines indicate the locations of the inner and outer cylinders.}
		\label{fig:Gene}
	\end{figure}

	The axisymmetric wave generator \citep{maurer2017}, adapted from a previous planar version \citep{gostiaux2006}, is used to produce inertial waves via a prescribed motion of ring-shaped elements discretizing the annular upper bound of the fluid domain. This device has been slightly modified to fit our needs, by lowering down the cylinders with $20\mathrm{~cm}$ long aluminium rods. In the configuration presented in figure~\ref{fig:Gene}, the five inner cylinders have been removed. The motion amplitudes of the remaining eleven outer cylinders (grey boxes in the cross-section in figure~\ref{fig:Cuve}(a)) have been adjusted to preserve the volume of fluid displaced during its motion, such that the profile $z(r)$ of the generator satisfies
	\begin{equation}
		\int_{R_0}^{R_1} z(r) r \diff r = 0.
	\end{equation}
	In order to preserve the boundary condition of non pumping fluid at the cylinder edges, the radial velocity $v_r$ has to be zero at $R_0$ and $R_1$. This condition writes, in terms of the profile $z(r)$, as
	\begin{equation}
		\dfrac{\diff z}{\diff r} (r = R_0) = \dfrac{\diff z}{\diff r} (r = R_1) = 0.
	\end{equation}	
	As shown in \cite{boury2019}, this facility is efficient in producing modes $1$ to $3$ Bessel shaped profiles, though the discretization of the wave generator leads to lower resolved modes at high order. We therefore looked for the closest approximation of a mode $1$ profile in such a confined geometry. The selected profile is a cubic shaped profile, as shown by the dashed line in figure~\ref{fig:Gene}, that sets the cam motion amplitudes. The highest amplitude for a cam is $a=2.5\mathrm{~mm}$ next to the inner cylinder, low enough to ensure a gradual growth of non-linear effects.

	The commonly used Particle Image Velocimetry (PIV) technique was implemented to visualise the velocity fields. Horizontal and vertical laser planes were created using a $2\mathrm{~W}$ Ti:Sapphire laser (wavelength $532\mathrm{~nm}$) and a cylindrical lens. While filling the tank, hollow glass spheres and/or silver coated spheres of $10\mathrm{~\mu m}$ diameter were added for the purpose of visualisation. Particle displacements were recorded at $40\mathrm{~Hz}$ using a camera located either on the side of the tank (vertical-plane visualisation) or down below facing a $45^\circ$ mirror placed under the tank (horizontal-plane visualisation). The CIVx algorithm was subsequently used to process the PIV raw images and extract the velocity fields \citep{fincham2000}.

\section{Experimental results}

	Before getting to the description of the experimental results, let us make a few notes on the geometry of the setup and the choice of the parameter range. As discussed in the introduction, the setup is designed to allow for the non-linear coupling between [i] inertial wave attractors in vertical (meridional) cross-sections and [ii] slow two-dimensional manifold in horizontal (equatorial) cross-section. We chose the geometric configuration of the conical bottom with the apex pointing upwards because, in such a geometry, the inertial waves undergo an additional focusing due to geometrical convergence of waves propagating from larger to smaller radial coordinate besides a primary focusing due to the reflection on the cone. This additional focusing favours the onset of instability close to the inner cylinder, as can be seen in the preliminary numerical study of \cite{Sibgatullinetal2017}. Furthermore, in the saturated regime we expect to localise the slow two-dimensional manifold in the vicinity of the inner cylinder, thereby facilitating the observation of a ``vortex condensate'' on top of ``inertial-wave gas'' background.

	To reduce the effect of viscosity on the non-linear energy transfer we chose a rather high rotation rate of the setup, $\Omega=2.093 \mathrm{~rad\cdot s^{-1}}$, so that the relevant value of the Ekman number in the present experiments is reasonably low $E=\nu /(2\Omega L^2) \approx 1.1 \cdot 10^{-5}$. The global {\it a priori} Rossby number based on the horizontal scale of the fluid domain, $Ro^{L}=U/(2\Omega L)$, can be defined using the maximum vertical speed of the generator rings as the velocity scale so that $U=a \omega_0$, where $\omega_0$ is the forcing frequency. For our experimental conditions, the amplitude $a=2.5\mathrm{~mm}$ and frequency $\omega_0=1.7\mathrm{~rad\cdot s^{-1}}$ yield $Ro^{L}= 7 \cdot 10^{-3}$ so that the system is expected to be strongly affected by the Coriolis force. This low value of the global Rossby number, however, corresponds to a developed non-linear energy cascade so that the observed saturated regime is more appropriately characterized by relevant micro-Rossby numbers discussed below.  

		\subsection{Linear regime}

In the linear regime we recover the classical dynamics: at the time-scale of order $10T_0$ after the start of the forcing, where $T_0=2\pi/\omega_0$ is the forcing period, iterative focusing downscales the wave motion from the global scale $L$ to the scale associated with the width of the wave beams \citep{Rieutordetal2001,HBDM2008,Grisouardetal2008}. Typical wave patterns observed in the quasi-linear regime at $t=17 T_0$ in horizontal and vertical planes are presented in the upper rows of figures~\ref{fig:pivnonlinearbis} and~\ref{fig:pivnonlinear} in terms of the quantities filtered at $\omega=\omega_0$ and $\omega=0$, respectively. For clarity, we visualise  the fields of radial $v_{r}$ and azimuthal $v_{\theta}$ velocity, and vertical vorticity $\xi_{z}$ in the horizontal (equatorial) plane, and the corresponding field of vertical velocity in the vertical (meridional) plane. It can be seen that the wave pattern observed at the forcing frequency (figure~\ref{fig:pivnonlinearbis}) in the  horizontal plane is to a good approximation axisymmetric, while in the vertical trapezoidal cross-section we recover a classic pattern of the $(1,1)$ wave attractor \citep{Maasetal1997, maas2001} in agreement with the ray tracing, whose branch width is due to an equilibrium between wave focusing and viscous dissipation~\citep{Rieutordetal2001,HBDM2008,Grisouardetal2008}. The experimental signal filtered around $\omega=0$ remains weak at $t=17T_0$ (see figure~\ref{fig:pivnonlinear}).

\subsection{Non-linear regime}

	\begin{figure}
	\centering
	\includegraphics[width=1\textwidth]{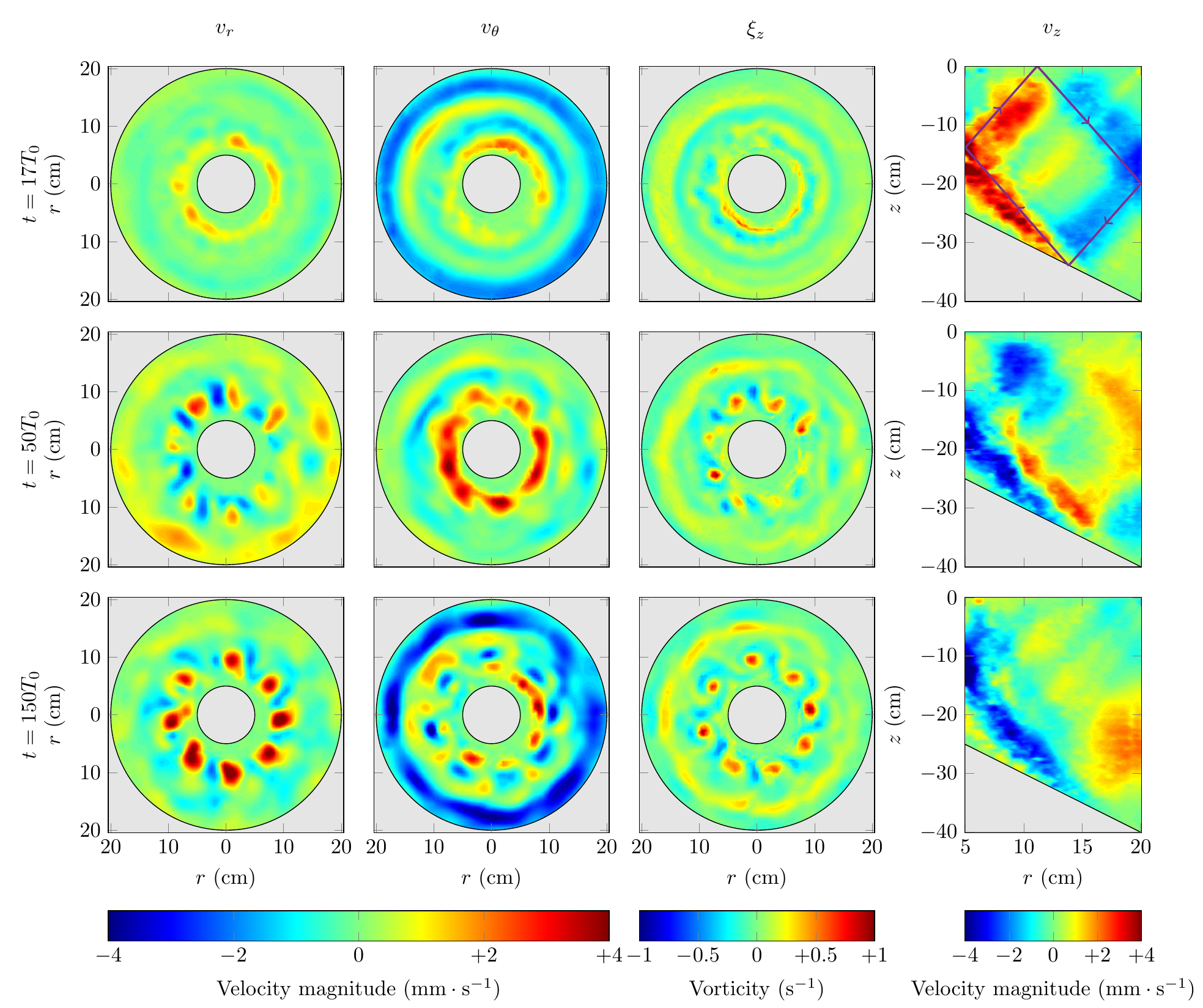}
	\caption{Fields of radial $v_r$ and azimuthal $v_{\theta}$ velocity and vertical vorticity $\xi_{z}$ in the horizontal plane (at $\simeq 20\mathrm{~cm}$ depth) and of vertical velocity $v_z$ in the vertical plane. The presented quantities are filtered around $\omega=\omega_0$. Positive vorticity corresponds to cyclonic motion. The parallelogram with arrows in the vertical plane shows the theoretical attractor in which the energy propagates clockwise.}
	\label{fig:pivnonlinearbis}
\end{figure}

\begin{figure}
	\centering
	\includegraphics[width=1\textwidth]{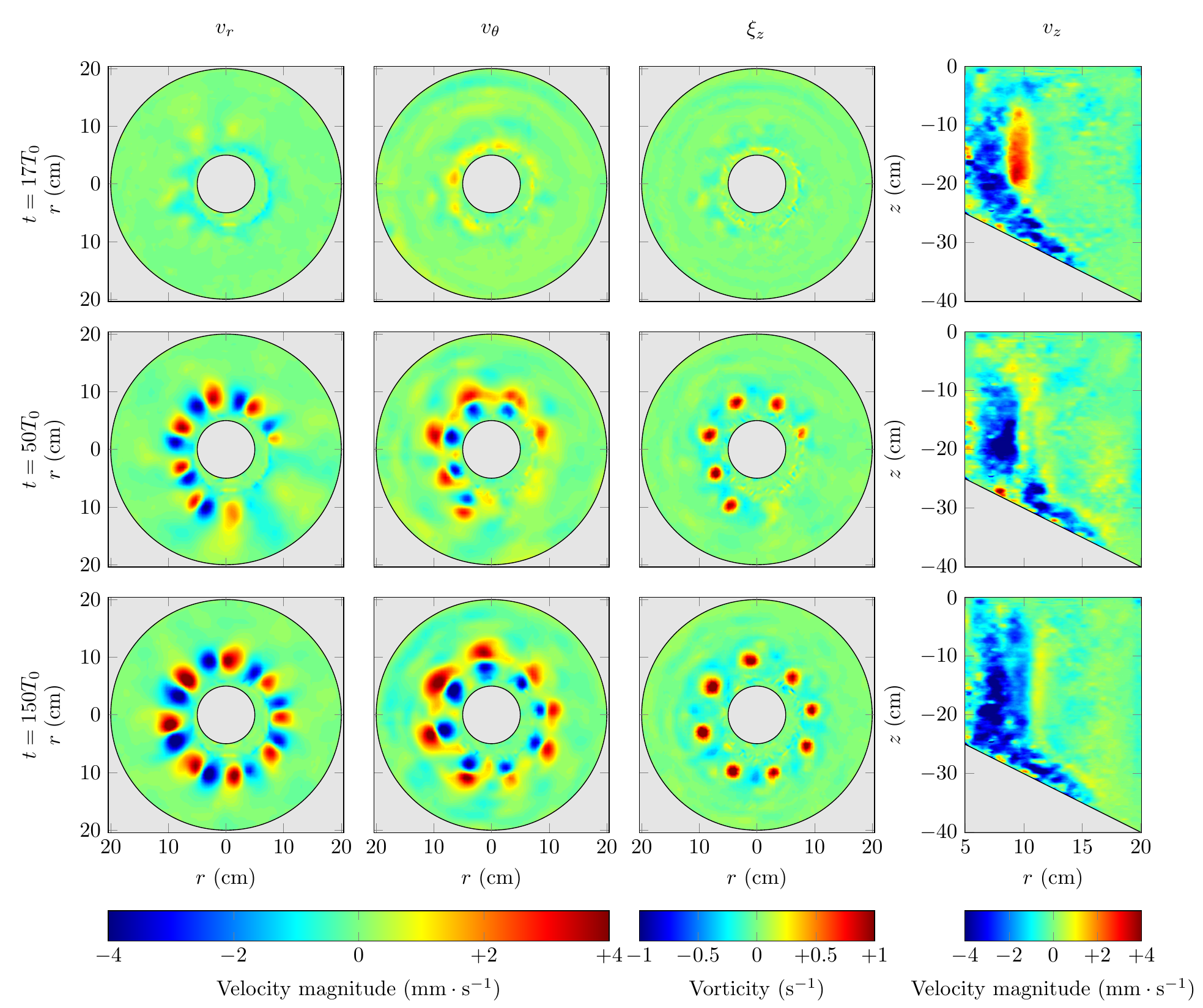}
	\caption{Fields of radial $v_r$ and azimuthal $v_{\theta}$ velocity and vertical vorticity $\xi_{z}$ in the horizontal plane (at $\simeq 20\mathrm{~cm}$ depth) and of vertical velocity $v_z$ in the vertical plane. The presented quantities are filtered around $\omega=0$. Positive vorticity corresponds to cyclonic motion.}
	\label{fig:pivnonlinear}
\end{figure}

	The development of the fully saturated non-linear regime is illustrated in figures~\ref{fig:pivnonlinearbis} and~\ref{fig:pivnonlinear} by snapshots corresponding to $t=50T_0$ and $t=150T_0$. The full vortex pattern representing the slow two-dimensional manifold is formed at the time scale of $100T_0$. It can be clearly seen that the initial axisymmetry observed at $t=17T_0$ is lost while the slow manifold is gradually formed. The latter is represented by a regular polygonal system of eight cyclonic vortices. The vortices are nearly invariant in the vertical direction as attested by the right column of images in figure~\ref{fig:pivnonlinear}, representing the vertical velocity component. These vortex structures are reminiscent of the Taylor columns usually found in rotating systems. There is, however, a crucially important distinction: while the Taylor columns are normally formed as a consequence of a slow motion of a perturbation imposed on the rotating fluid, the coherent structures seen in figure~\ref{fig:pivnonlinear} arise due to a non-linear process which drains energy from the wave field toward the slow two-dimensional manifold. The vertical velocity in cyclonic vortices is directed downwards, corresponding to Ekman pumping, in agreement with existing experimental and numerical data (e.g. \cite{Hopfingeretal1982,GodeferdLollini1999}). It is worth mentioning that even in the fully saturated regime one can still identify the branches of the inertial wave attractor in the signal filtered around the forcing frequency $\omega=\omega_0$ (see figure~\ref{fig:pivnonlinearbis}). The relevance of such experimental regime (where a wave attractor in the meridional plane co-exists with a polygonal vortex system in the equatorial plane) to geo- and astrophysical systems admitting the existence of inertial-wave attractors \citep{dintrans1999,RieutordetalPRL2000,RieutordValdettaro2010} represents an interesting direction for future research.

 

\begin{figure}
		\centering
		\includegraphics[width=1\textwidth]{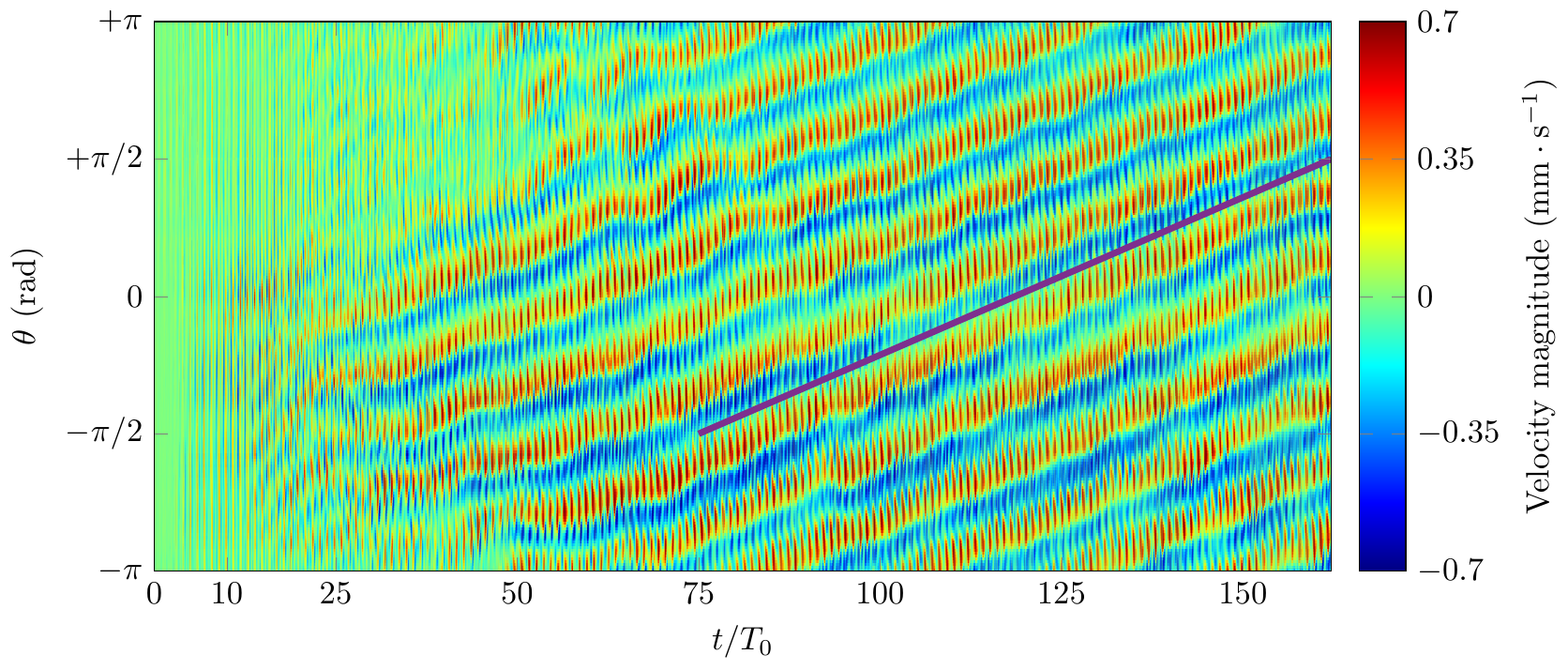}
		\caption{Temporal evolution of the radial velocity as a function of azimuth, $\theta$, obtained by taking profiles of radial velocity at radius $r= 8\mathrm{~cm}$ in the horizontal plane at $z\simeq 20\mathrm{~cm}$ above the bottom. To guide the eye, we added a solid line showing that the cluster rotates half a turn ($\pi~\mathrm{rad}$) in $87 T_0$.}
		\label{fig:radevol}
	\end{figure}

	The visual evidence of the transients is seen in figure~\ref{fig:radevol} which presents the temporal evolution of the radial structure measured by sampling the azimuthal distribution of radial velocity at the radius $r= 8\mathrm{~cm}$ corresponding to the position of centres of vortices seen in figure~\ref{fig:pivnonlinear}. Figure~\ref{fig:radevol} shows that coherent structures start to appear after roughly $25T_0$, and that further the vortex pattern self-organises itself so that new structures gradually appear and join the ensemble. After roughly $100T_0$ all structures move at the same rate in cyclonic direction. This rate corresponds to one complete turn of the vortex cluster around the inner cylinder in about $180T_0$.

\begin{figure}
		\centering
		\includegraphics[width=0.9\textwidth]{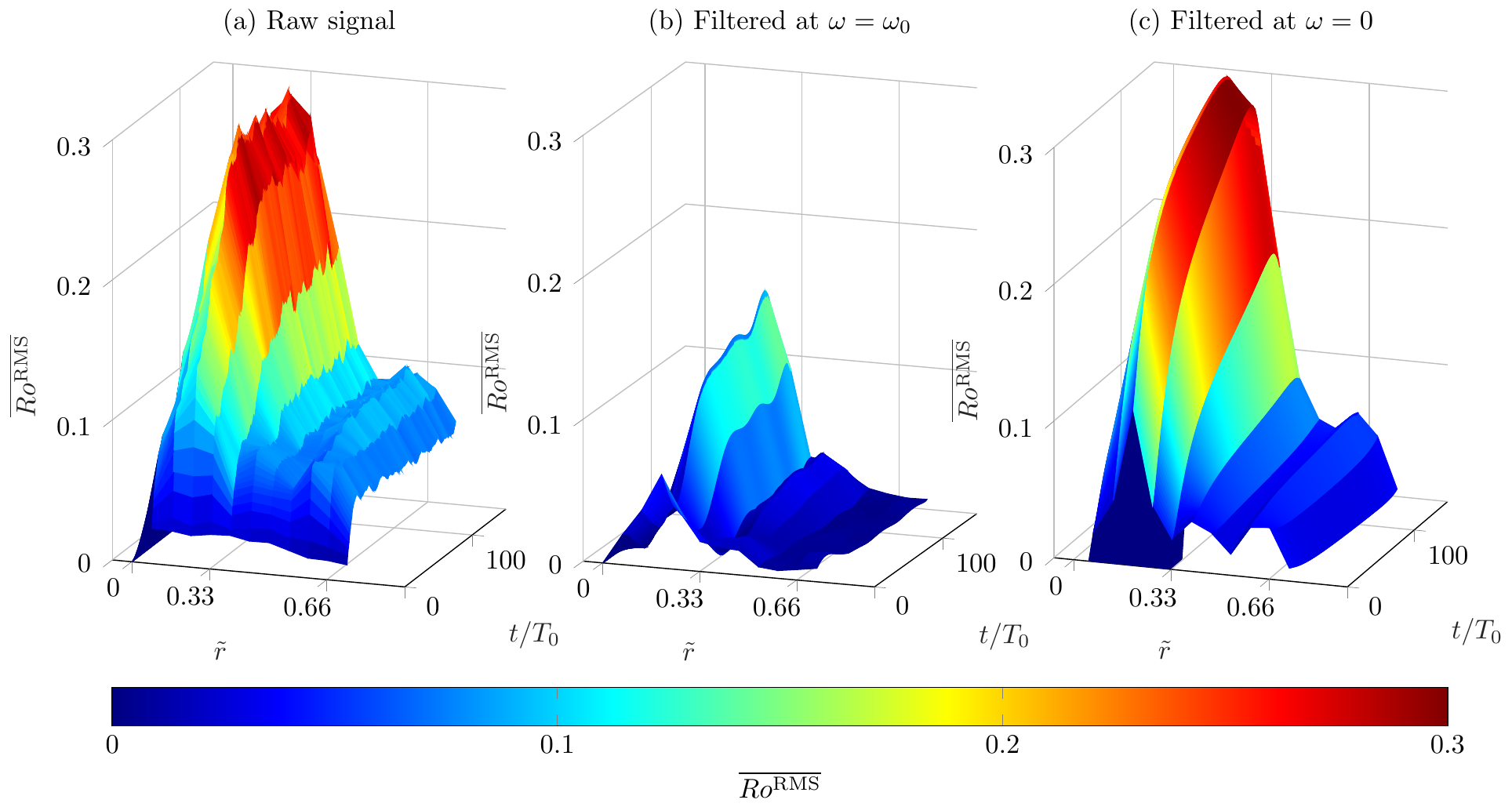}
		\caption{Spatio-temporal surface plot of the zonally- and period-averaged Rossby number. The curves (from left to right) correspond to processing of the raw signal, and the signal filtered around $\omega=\omega_0$ and $\omega=0$.}
		\label{fig:surfrossby}
	\end{figure}

	For better insight into the dynamics of the process, let us consider the time-evolution of the zonally- and period-averaged Rossby number $\overline{Ro^\mathrm{RMS}}=\overline{\left\langle\xi_{z}\right\rangle}/(2\Omega)$. To do this we split the distance $L$ between the inner and outer cylinders into $M=10$ equally spaced ring-shaped zones, and calculate the RMS value of vertical vorticity $\xi_{z}$ over the area of these zones. This operation is denoted by brackets $\left\langle \cdot \right\rangle$. Further, we calculate the period-averaged quantity denoted by overbar. These operations are performed for the raw signal of vertical vorticity component $\xi_{z}$, and the values filtered around the forcing $\omega=\omega_0$ and zero $\omega=0$ frequencies. The evolution of the averaged Rossby number as function of radial coordinate and time  $\overline{Ro^\mathrm{RMS}}(\tilde{r},t/T_0)$, where $\tilde{r}=(r-R_0)/(R_1-R_0)$, is shown in figure~\ref{fig:surfrossby}. It can be seen that after $t=100T_0$ one obtains roughly time-invariant saturated profiles $\overline{Ro^\mathrm{RMS}}(\tilde{r},\infty)$. It can be seen that the main contribution to the value of $\overline{Ro^\mathrm{RMS}}$ calculated over the raw signal is due to the zero-frequency component. The maximum value of the Rossby number $\overline{Ro^\mathrm{RMS}}$ in the saturated regime is about $\overline{Ro^\mathrm{RMS}_\mathrm{max}} \approx 0.28$, and this value is reached at {$\tilde{r} \approx 0.25$}, which corresponds to the distance where the centers of vortex structures are located. Let us note that the numerical simulations presented in \cite{BourouibaBartello2007} suggest that at $Ro^\mathrm{RMS}=0.2$ the three-dimensional to two-dimensional energy transfer is maximized. Our measurements performed in a quite different system yield a comparable value. 
	
\begin{figure}
		\centering
		\includegraphics[width=1\textwidth]{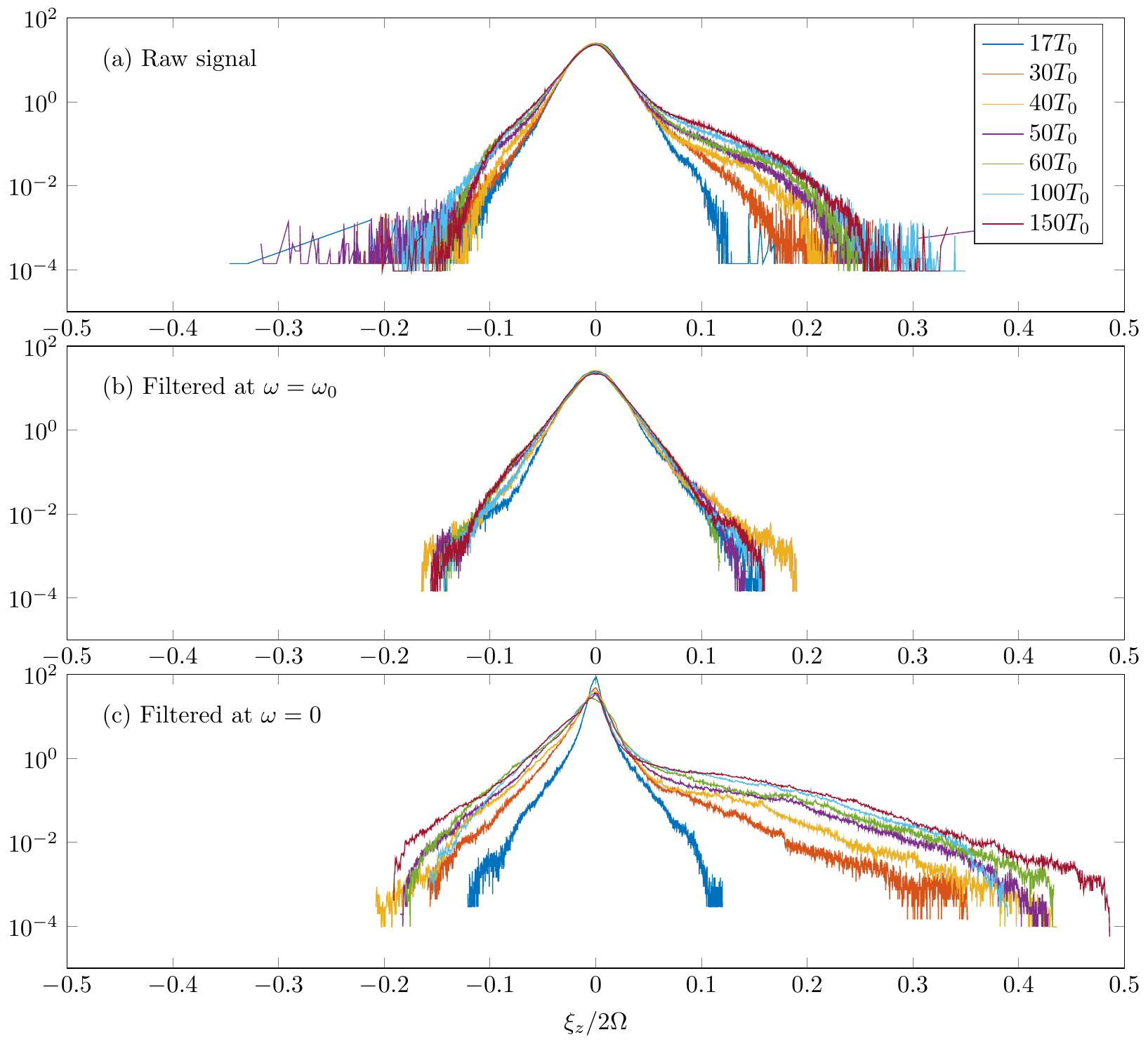}
		\caption{PDFs of the vertical vorticity component at different times in the experiment. The curves (from top to bottom) correspond to processing of the raw signal, and the signal filtered around $\omega=\omega_0$ and $\omega=0$.}
		\label{fig:pdfs}
	\end{figure}

	In addition to the evolution of $\overline{Ro^\mathrm{RMS}}$ in space and time, the vertical vorticity field in the horizontal plane can be characterized statistically, by measuring the Probability Density Function (PDF) of the micro-Rossby number $Ro^{\xi_{z}}=\xi_{z}/(2\Omega)$. The typical PDFs of $Ro^{\xi_{z}}$ corresponding to different stages of development of the coherent vortex structures are shown in figure~\ref{fig:pdfs}. Similar to the data presented in figure~\ref{fig:surfrossby}, we calculate the PDFs for the raw signal and for the signal filtered around the forcing $\omega=\omega_0$ and $\omega=0$ frequencies. The PDFs are calculated over the surface of the ring-shaped zone between the inner and outer cylinders, and over the time-span of $\pm2T_0$ around the time instances indicated in figure~\ref{fig:pdfs}. It can be seen that at the beginning of the process, when the motion is represented essentially by the axisymmetric waves, the PDFs of $Ro^{\xi_z}$ has a sharp symmetrical form. As the non-linear energy transfer towards coherent vortex structures develops, there is a progressive evolution of the vorticity PDFs toward the shape characterized by asymmetric ``shoulders'', which indicates a well-pronounced cyclonic/anticyclonic asymmetry. This asymmetry is clearly seen in the PDFs calculated over the raw signal and the signal filtered around zero frequency $\omega=0$ (upper and lower images in figure~\ref{fig:pdfs}), suggesting that a few strong cyclonic vortices seen in figure~\ref{fig:pivnonlinear} are responsible for the asymmetry of the PDFs. The probability density functions calculated for the signal filtered at the forcing frequency (middle image in figure~\ref{fig:pdfs}) remain approximately symmetrical at all time. A slight asymmetry visible in the curve corresponding to time around $t=30T_0$ and $40T_0$ can be tentatively attributed to the process of genesis of the regular polygonal vortex pattern: new cyclonic vortices are emerging in the plane of visualization and are joining the ensemble. Nonetheless, since the asymmetry is weak one cannot exclude also a minor contribution from experimental noise. 

	The observed energy cascade could be described by a tentative scenario that includes [i] energy injection at large scale by axisymmetric forcing, [ii] energy focusing at attractor triggering [iii] a cascade of triadic interactions generating secondary waves with nearly vertical group velocity vector, [iv] energy drain toward cyclonic vortices via wave-vortex interactions, [v] self organization of a regular polygonal vortex pattern stabilized by Ekman pumping in a saturated regime.

\section{Conclusions}

	Investigation of inertial wave attractors in rotating fluids offer a number of possibilities concerning the geometric setup of the problem. A number of experimental studies \citep{maas2001, MandersMaas2003,MandersMaas2004,brunet2019} consider a rectangular box geometry with one sloping wall placed at a rotating table which is similar to the geometry used to reveal the linear \citep{Maasetal1997} and non-linear \citep{Scolanetal2013,BrouzetetalEPL2016,Brouzetetal2017,Davisetal2020} dynamics of internal wave attractors in stratified fluids. It has been realized that the inherent three-dimensionality of inertial waves is responsible for considerable secondary currents \citep{maas2001,MandersMaas2004} and for a number of notable changes in the scenario of Triadic Resonance Instability \citep{maurer2016, brunet2019}. On the other hand, there is a rich theoretical literature which considers linear viscous regimes of inertial wave attractors in spherical liquid shells, where the flow is studied in the meridional cross-section while the structure of the flow in the equatorial cross-section is supposed to be trivial \citep{RieutordValdettaro1997,RieutordetalPRL2000,RieutordValdettaro2010}. 

	In the present paper, we consider the experimental setup which builds a bridge between the two above statements: the experiments are carried out in a (horizontal) annular and (vertical) trapezoidal domain which admits the existence of inertial wave attractor structures in meridional planes while leaving the freedom for the formation of a slow two-dimensional manifold which drains energy from the genuinely three-dimensional inertial wave field as result of the energy cascade. The experimental system is subject to axisymmetric forcing, and experiments are performed at low values of the global Rossby ($Ro$ is of order $10^{-3}$) and Ekman numbers ($E$ is of order $10^{-5}$). The main finding of this study is the non-trivial formation of a two-dimensional manifold in the saturated regime in the equatorial plane, which co-exists with an inertial wave attractor in the meridional plane. The two-dimensional manifold is represented by an ensemble of eight cyclonic vortices in a regular polygonal arrangement. The vortex system as a whole undergoes a slow cyclonic motion around the axis of rotation of the experimental system. The probability density function of the vertical vorticity has asymmetric ``shoulders'' which are indicative of cyclonic/antisyclonic asymmetry, a well-known property of rotating turbulence. The main contribution to the PDFs asymmetry in our experimental system is due to the concentrated cyclonic vortices.

	The findings presented in this study raise a number of interesting issues, in particular, [i] how the saturated vortex regime depends on the key parameters of the problem (Rossby and Ekman numbers, geometric aspect ratio and the particular type of forcing), [ii] whether or not the observed regime might be relevant to realistic geo- and astrophysical systems. Cyclonic clusters arranged in form of regular polygones have been reported for the polar regions of large planets, e.g. Jupiter \citep{Adrianietal2018}, demonstrating remarkably persistent long-term behavior \citep{Adrianietal2020}. It has been argued that although the global pattern of the Jovian atmosphere with circumpolar cyclones and alternating flows near the equator is captured by shallow-water models (e.g. \cite{ChoPolvani1996,ScottPolvani2007}) the key puzzle remains \citep{Adrianietal2018}: ``The manner in which the cyclones persist without merging and the process by which they evolve to their current configuration are unknown''. Recently, \cite{reinaud2019} has shown numerically that a system of $m$ quasi-geostrophic vortices equally distributed over a ring whose center is already occupied by a vortex can be stable under certain conditions. In our experiment, the inner cylinder may play a similar role to this central vortex. Hence, the ``toy system'' proposed here may help to shed light on the issues related to the stability of polygonal vortex clusters and their possible existence in the interiors of natural systems admitting the existence of inertial wave attractors.

\bigskip
\textbf{Acknowledgments}

This work was supported by the grant ANR-17-CE30-0003 (DisET) and by the LABEX iMUST (ANR-10-LABX-0064) of Universit\'e de Lyon, within the program “Investissements d'Avenir” (ANR-11-IDEX-0007), operated by the French National Research Agency (ANR). This work was also supported by a grant from the Simons Foundation (651475, TD). It has been achieved thanks to the resources of PSMN from ENS de Lyon. EE gratefully acknowledges his appointment as a visiting scientist at ENS de Lyon during the experimental campaign. EE and NS acknowledge support from the Russian Science Foundation (Project 20-11-20189) during work with the manuscript and complementary data processing. 

\bigskip
Declaration of Interests. The authors report no conflict of interest.

\bibliographystyle{jfm}
\bibliography{bibIWA}

\end{document}